\documentclass[prd,twocolumn,showpacs,floatfix,amsmath,amssymb,floatfix]{revtex4}
\usepackage{graphicx,color,dcolumn,booktabs,bm}
\usepackage{longtable,lscape}
\usepackage{txfonts}
\usepackage{overpic}
\usepackage{amssymb}
\usepackage{indentfirst}
\usepackage{feynmf}   
\usepackage{slashed}  
\usepackage{cases}
\usepackage{color}
\usepackage{multirow}
\usepackage{epstopdf}
\usepackage{graphicx,color,dcolumn,booktabs,bm}

\graphicspath{{Figures/}} %
\usepackage[colorlinks,
                      citecolor=blue,
                      anchorcolor=red,
                      menucolor=red,
                      linkcolor=red,
                      filecolor=red,
                      runcolor=red,
                      urlcolor=blue,
                      frenchlinks=red]{hyperref}
\begin{document}

\title{The decays of the fully open flavor state $T_{c\bar{s}0}^{0}$ in a $D^{*}K^{*}$ molecule scenario}
\author{Zi-Li Yue$^{1}$}
\author{Cheng-Jian Xiao$^2$}
\author{Dian-Yong Chen$^{1,3}$\footnote{Corresponding author}} \email{chendy@seu.edu.cn}
\affiliation{
 $^{1}$ School of Physics, Southeast University,  Nanjing 210094, China}
\affiliation{
 $^{2}$ Institute of Applied Physics and Computational Mathematics, Beijing 100088, China} 
\affiliation{$^3$Lanzhou Center for Theoretical Physics, Lanzhou University, Lanzhou 730000, China}
\begin{abstract}
Inspired by the recent observations of $T_{c\bar{s}0}^{0/++}$ in the the processes $B^0\to\bar{D}^0 D_s^+ \pi^-$ and $B^+\to D^- D_s^+ \pi^+$ by LHCb Collaboration, we investigate the decay properties of the $T_{c\bar{s}0}^{0}$ in a $D^{*}K^{*}$ molecule scenario, and the widths of $T_{c\bar{s}0}^{0}\to D^{0}K^{0}$,~$D_{s}^{+}\pi^{-}$,~$D_{s}^{*+}\rho^{-}$,~$D_{s1}^{(\prime)+}\pi^{-}$, and $D^{*0}(D\pi)^{0}$ are estimated. Our estimations indicate that the width of  $T_{c\bar{s}0}^{0} \to D_s^+ \pi^-$ is sizable to be observed and the dominant decay mode of $T_{c\bar{s}0}^{0}$ is $D^0K^0$. Considering the isospin symmetry, we proposed to search $T_{c\bar{s}0}(2900)^{++}$ in the $D^+ K^+$ invariant mass distributions of the process $B^+ \to D^+ D^- K^+$, where some preliminary experimental hints have been observed by LHCb Collaboration.
\end{abstract}


\maketitle

\section{Introduction}
\label{sec:introduction}

Searching for the fully open flavor tetraquark states are particular interesting since their quark components are obviously different with the traditional mesons, which makes them much easier to be identified as a tetraquark states. The first instance of hadronic state with valence quarks of four different flavors, $X(5568)$, was reported in the mass spectrum of $B_s^0 \pi^\pm$ based on $10.4~\mathrm{fb}^{-1}$ of $p\bar{p}$ collision data by the D0 experiment at the Fermilab Tevatron collider in the year of 2016~\cite{D0:2016mwd, D0:2017qqm}. After the observation of $X(5568)$ by the D0 Collaboration, the LHCb~\cite{LHCb:2016dxl}, CMS~\cite{CMS:2017hfy},  CDF~\cite{CDF:2017dwr}, and ATLAS~\cite{ATLAS:2018udc} Collaborations searched the $X(5568)$ in the same final states successively, but no signal was observed. On the theoretical side, there are different views in the existence of $X(5568)$, for example, almost all the investigations in Refs.~\cite{Agaev:2016mjb,Zanetti:2016wjn, Chen:2016mqt, Wang:2016mee, Liu:2016ogz, Agaev:2016lkl, Agaev:2016ijz, Wang:2016wkj, Dias:2016dme, Wang:2016tsi, Agaev:2016urs, Liu:2016xly, Jin:2016cpv, He:2016yhd, Lu:2016zhe, Burns:2016gvy, Xiao:2016mho} supported the existence of $X(5568)$ and could be considered as a $B K$ molecular state or $\bar{b}s \bar{q}q$ compact tetraquark state, while the authors in Refs.~\cite{Guo:2016nhb,Albaladejo:2016eps, Chen:2016npt, Lang:2016jpk, Kang:2016zmv, Chen:2016ypj, Lu:2016kxm, Goerke:2016hxf, Agaev:2016ifn} had just the opposite opinions.


Another two additional fully open flavor tetraquark states are $X_{0}(2900)$ and $X_1(2900)$, which were first observed in the $D^{-}K^{+}$ invariant mass spectrum of the process $B^{+}\to D^{+}D^{-}K^{+}$ in 2020~\cite{LHCb:2020bls,LHCb:2020pxc}. Their most possible quark components are $ud\bar{c}\bar{s}$, and the fully open flavor property of these two states has attracted the theorists' great interest and various interpretations have been proposed by different groups. The estimations in the constituent quark model~\cite{Karliner:2020vsi} and QCD Sum Rule~\cite{Zhang:2020oze,Wang:2020xyc} indicate that $X_0(2900)$ could be a compact tetraquark state with $I(J^P)=0(0^+)$, while the estimations in Ref.~\cite{Mutuk:2020igv} supported the diquark-antidiquark tetraquark interpretation for $X_1(2900)$, but for $X_0(2900)$, they found it could be a $D^\ast K^\ast $ molecular state. The estimations of   QCD two-point sum rule method~\cite{Agaev:2020nrc} and potential model~\cite{Molina:2020hde, Xue:2020vtq, Liu:2020nil,Chen:2020aos,Huang:2020ptc,He:2020btl} supported $X_0(2900)$ as $D^\ast \bar{K}^\ast$ molecular state. Moreover, in Ref.~\cite{Xiao:2020ltm}, the authors investigated the decay properties of the $X_{0}(2900)$ in the $D^\ast K^\ast $ molecular scenario based on an effective Lagrangian approach.


Very recently, the LHCb collaboration reports two resonances $T_{c\bar{s}0}(2900)^{0/++}$ (abbreviate to $T_{c\bar{s}0}^{0/++}$ here and after), which are two of the isospin triplet, in the $D_{s}\pi$ invariant mass spectrum of the processes $B^0 \to \bar{D}^0 D_s^+ \pi^-$ and $B^+ \to D^- D_s^+ \pi^+$ with a significance to be $9\sigma$. The analysis indicated that the spin-parity is preferred to be $J^P=0^+$. The resonance parameters of $T_{c\bar{s}0}^{0/++}$ are measured to be~\cite{LHCb:New1, LHCb:New2},
\begin{eqnarray}
m_{T_{c\bar{s}0}^{0}}&=&(2892\pm14\pm15)~\mathrm{MeV},\nonumber\\
\Gamma_{T_{c\bar{s}0}^{0}}&=&(119\pm26\pm12)~\mathrm{MeV},
\end{eqnarray}
and
\begin{eqnarray}
m_{T_{c\bar{s}0}^{++}}&=&(2921\pm17\pm19)~\mathrm{MeV},\nonumber\\
\Gamma_{T_{c\bar{s}0}^{++}}&=&(137\pm32\pm14)~\mathrm{MeV},
\end{eqnarray}
respectively. The quark components of $T_{c\bar{s}0}^0$ are $c\bar{s} \bar{u} d$, which is also a fully open flavor state. The estimations in the constituent quark model~\cite{Liu:2022hbk} and multiquark color flux-tube model~\cite{Wei:2022wtr} suggested that $T_{c\bar{s}0}$ can be assigned to be a tetraquark state. While the authors in Ref.~\cite{Molina:2022jcd} interpreted the $T_{c\bar{s}0}$ as a threshold effect from the interaction of the $D^*K^*$ and $D^*_s\rho$ channels. In addition, the observed mass of $T_{c\bar{s}0}$ is very close to the threshold of $D^\ast K^\ast$, which indicate that $T_{c\bar{s}0}$ could be a good candidate of molecular state composed of $D^\ast$ and $K^\ast$. In Ref.~\cite{Chen:2022svh}, the interactions between $D^{(\ast)}K^{(\ast)}$ were investigated by using the one-boson-exchange model and the estimation indicate that the $T_{c\bar{s}0}$ could be interpreted as a $D^\ast K^\ast$ molecular state with $I(J^P)=1(0^+)$. Along this way, in the present work we further inspect the possibility of the  $D^\ast K^\ast$ molecular interpretation by investigating the decay properties of $T_{c\bar{s}0}$, and try to find its dominant decay modes, which may be helpful for searching $T_{c\bar{s}0}$ in further experiments at the  Belle and LHCb Collaborations.

This work is organized as follows. The hadronic molecular structures of $T_{c\bar{s}0}$ is discussed in the following section and the possible decay channel, including the two-body and three body decays are estimated in section \ref{sec:Sec3}. The numerical results and the relevant discussions are presented in section \ref{sec:Sec4}, and the last section is dedicated to a short summary.

\section{Hadronic molecular structure}
\label{sec:Sec2}

\begin{figure}[t]
  \centering
  \includegraphics[width=7cm]{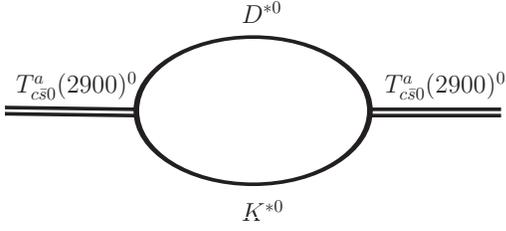}
  \caption{The mass operator of the $T_{c\bar{s}0}^{0}$.}\label{Fig:Tri1}
\end{figure}

In the present work, we consider $T_{c\bar{s}0}$ as a molecular state composed of $D^\ast K^\ast$ with $I(J^P)=1(0^+)$, and here we take the neutral one, $T_{c\bar{s}0}^0$, as an example. The effective Lagrangian between $T_{c\bar{s}0}^{0}$ and its components is,
\begin{eqnarray}
\mathcal{L}=g_{T}T_{c\bar{s}0}^{0}(x)\int dy\Phi(y^{2})D^{*0\mu}(x+\omega_{K^{*0}}y)K^{*0}_{\mu}(x-\omega_{D^{*0}}y)\label{Eq:1}.
\end{eqnarray}
where $\omega_{K^{*0}}=m_{K^{*0}}/(m_{K^{*0}}+m_{D^{*0}})$,~$\omega_{D^{*0}}=m_{D^{*0}}/(m_{K^{*0}}+m_{D^{*0}})$. $\Phi(y^{2})$ is the correlation function, which is introduced to describe the inner distributions of the component. The Fourier transformation of $\Phi(y^{2})$ is,
\begin{eqnarray}
\Phi(y^{2})=\int \frac{d^{4}p}{(2\pi)^{4}}e^{-ipy}\tilde{\Phi}(-p^{2}).\label{Eq:2}
\end{eqnarray}
the choices of the $\tilde{\Phi}(-p^{2})$ should satisfy the conditions that can describe the interior structure of the molecular state and fall fast enough in the ultraviolet region. Here we use the correlation function in the Gaussian form, which is~\cite{Faessler:2007gv, Faessler:2007us, Xiao:2020alj, Xiao:2019mvs, Xiao:2016hoa, Chen:2015igx},
\begin{eqnarray}
\tilde{\Phi}(p_{E}^{2})=\mathrm{exp}(-p_{E}^{2}/\Lambda_{T}^{2}),
\end{eqnarray}
where $\Lambda_{T}$ is a model parameter related to the distribution of components in the molecular state, $P_{E}$ is the Jacobi momentum used to describe the relative motion.

The coupling constant $g_{T}$ introduced in Eq.~(\ref{Eq:1}) could be determined by the compositeness condition, which means that the possibility of $T_{c\bar{s}0}^{0}$ to be a bare elementary state is zero~\cite{Weinberg:1962hj,Salam:1962ap}, i.e.,
\begin{eqnarray}
Z=1-\frac{\partial\Pi(m_{T_{c\bar{s}0}^{0}}^{2})}{\partial m_{T_{c\bar{s}0}^{0}}^{2}} =0.\label{Eq:3}
\end{eqnarray}
Based on the effective Lagrangian in Eq.~(\ref{Eq:1}), the particular form of mass operator corresponding to Fig.~\ref{Fig:Tri1} can be written as,
\begin{eqnarray}
\Pi ({m_{T_{c\bar{s}0}^{0}}^{2}})&=&\int \frac{d^{4}q}{(2\pi)^{4}}\tilde{\Phi}^{2}[-(q-\omega_{D^{*}}p)^{2},\Lambda_{T}^{2}]\nonumber\\&\times&\frac{-g^{\mu\nu}+q^{\mu}q^{\nu}/m_{D^{*}}^{2}}{q^{2}-m_{D^{*}}^{2}}\nonumber\\&\times&\frac{-g^{\mu\nu}+(p^{\mu}-q^{\mu})(p^{\nu}-q^{\nu})/m_{{K}^{*}}^{2}}{(p-q)^{2}-m_{K^{*}}^{2}}.\label{Eq:MO}
\end{eqnarray}
With Eqs. (\ref{Eq:3}) and (\ref{Eq:MO}), one can estimate the coupling constant $g_T$ depending on the model parameter $\Lambda_T$.

\section{Strong decays of $T_{c\bar{s}0}^{0}$}
\label{sec:Sec3}

\begin{figure}[t]
\begin{tabular}{cc}
  \centering
  \includegraphics[width=4.2cm]{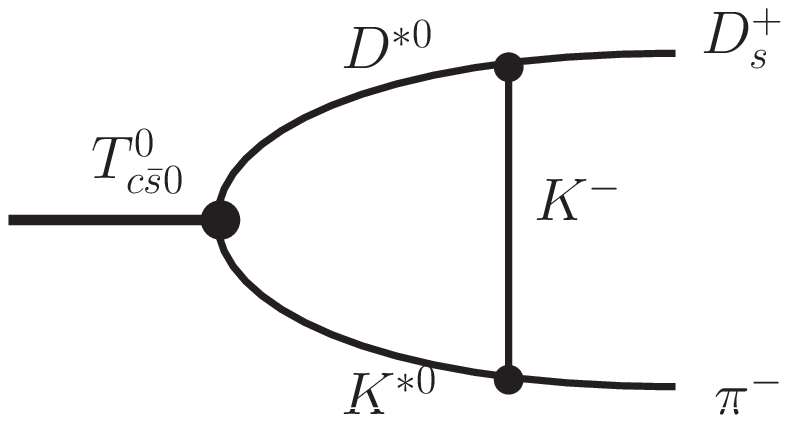}&
 \includegraphics[width=4.2cm]{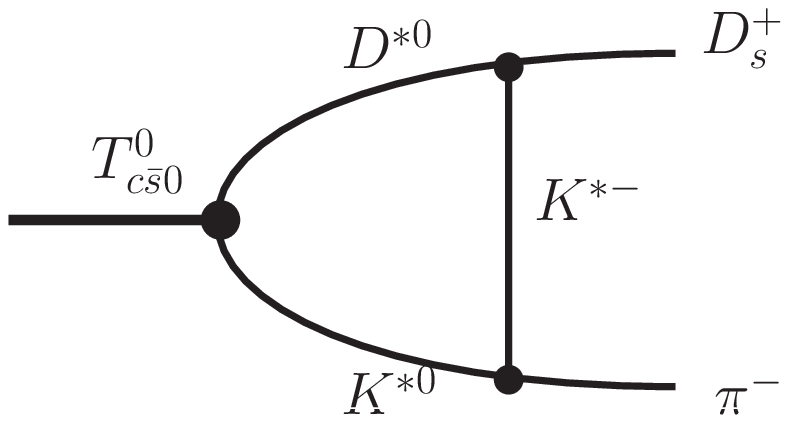}\\
\\
 $(a)$ & $(b)$ \\
  \includegraphics[width=4.2cm]{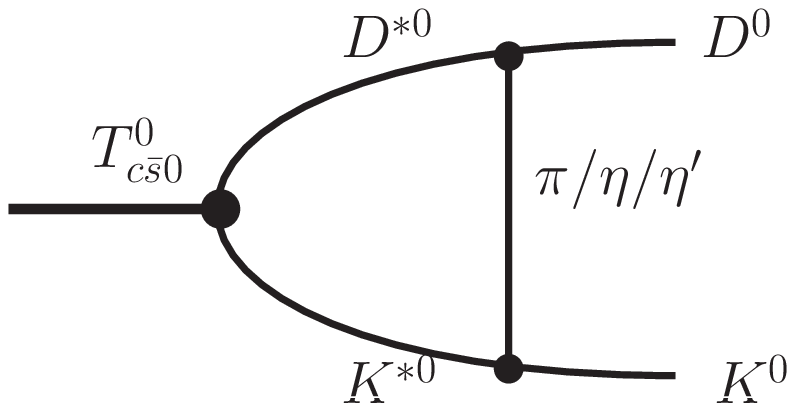}&
 \includegraphics[width=4.2cm]{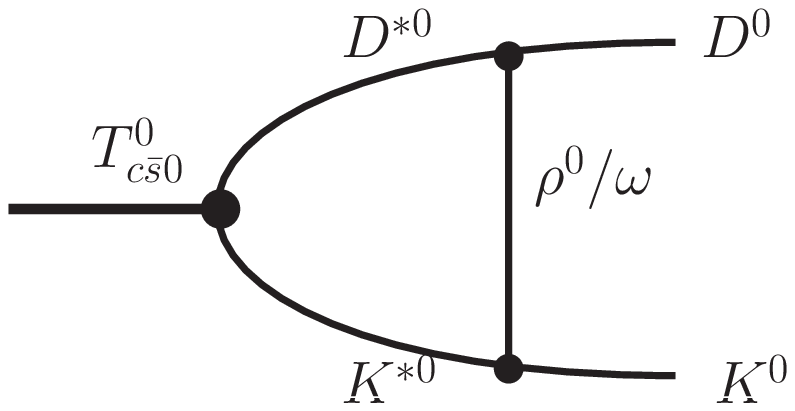}\\
\\$(c)$ & $(d)$ \\
\includegraphics[width=4.2cm]{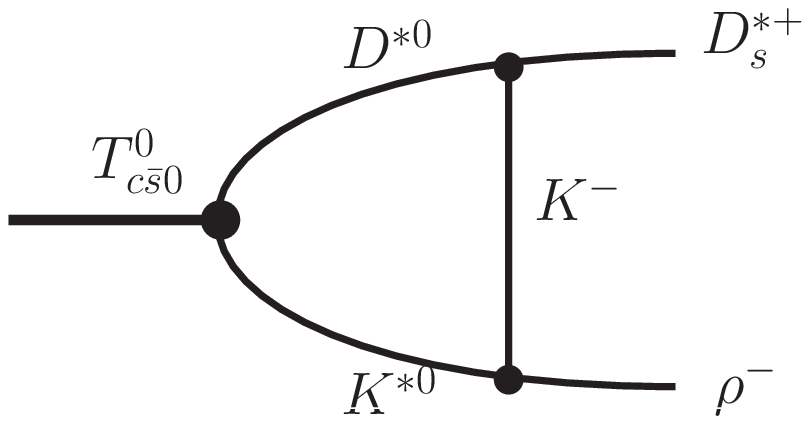}&
 \includegraphics[width=4.2cm]{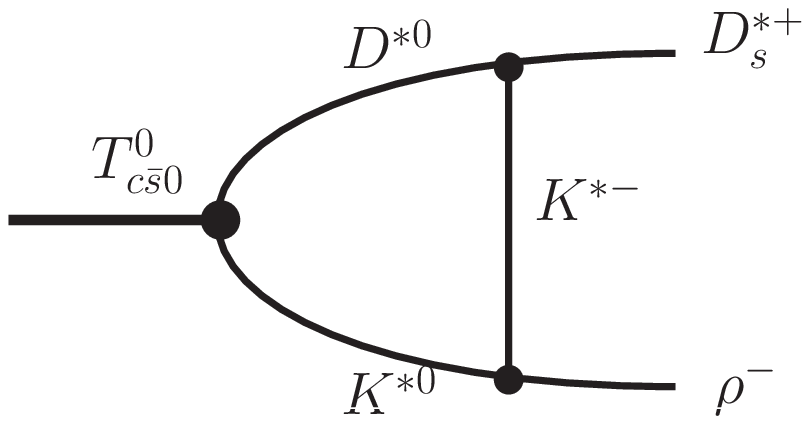}\\
\\$(e)$ & $(f)$ \\
\includegraphics[width=4.2cm]{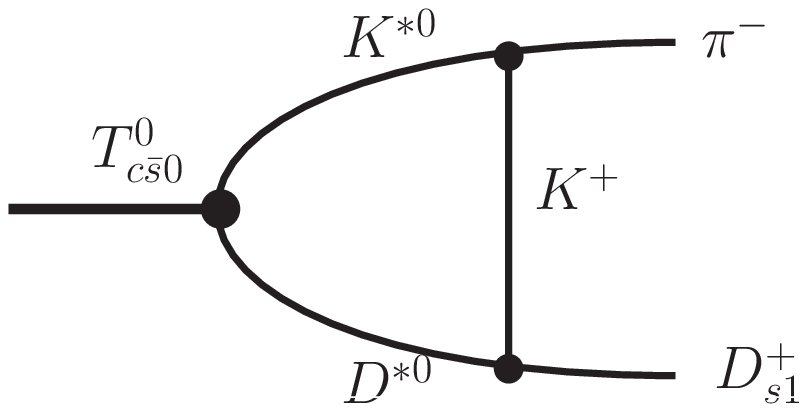}&
 \includegraphics[width=4.2cm]{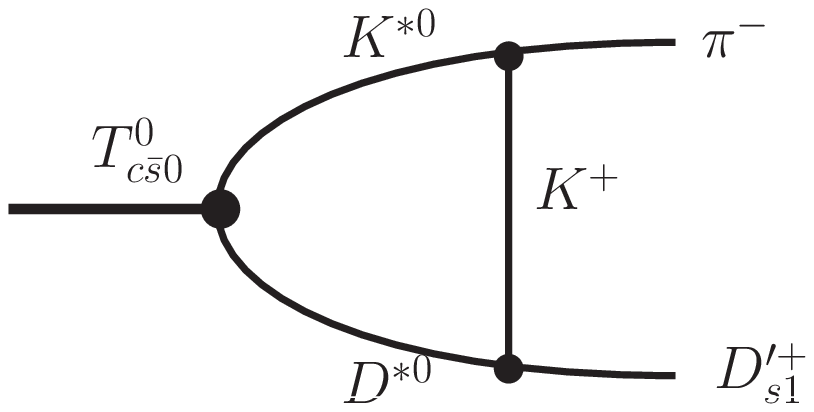}\\
\\$(g)$ & $(h)$ \\
 \end{tabular}
  \caption{The hadron level typical diagrams contributing to 
$T_{c\bar{s}0}^{0}\to D_{s}^{+}\pi^{-}$ (diagrams (a) and (b)), $T_{c\bar{s}0}^{0}\to D^{0}K^{0}$ (diagrams (c) and (d)), $T_{c\bar{s}0}^{0}\to D_{s}^{*+}\rho^{-}$ (diagrams (e) and (f)), $T_{c\bar{s}0}^{0}\to D_{s1}^{(\prime)+}\pi^{-}$ (diagrams (g) and (h)), the $D_{s1}$ and $D_{s1}^{\prime}$ refer to $D_{s1}(2460)$ and $D_{s1}(2536)$ respectively.}\label{Fig:Tri2}
\end{figure}

In the present work, we further inspect the possibility of $T_{c\bar{s}}$ as a $D^\ast K^\ast$ molecular state by investigating the decay behavior of $T_{c\bar{s}}$. As for $T_{c\bar{s}0}^0$, we find that the possible two body  decay channels include $T_{c\bar{s}0}^{0}\to D^{0}K^{0}$,~$D_{s}^{+}\pi^{-}$,~$D_{s}^{*+}\rho^{-}$,~$D_{s1}^{(\prime)+}\pi^{-}$, and three body decay channel $T_{c\bar{s}0}^0\to D^{\ast 0}(K\pi)^0$. The typical diagrams related to these decay processes are collected in Fig.~\ref{Fig:Tri2} and ~\ref{Fig:Tri3}.

\subsection{Effective Lagrangians}
In the present work, we estimate the decay process in the hadron level and the interaction between hadrons are described by effective Lagrangians. Considering the heavy quark limit and chiral symmetry, the relevant phenomenological Lagrangians are~\cite{Casalbuoni:1996pg,Oh:2000qr,Colangelo:2002mj,Kaplan:2005es,Kaymakcalan:1983qq},
\begin{eqnarray}
\mathcal{L}_{{\mathcal{D}}^{*} \mathcal{D} \mathcal{P}}&=&i g_{{\mathcal{D}}^{*} \mathcal{D} \mathcal{P}}\left({\mathcal{D}}^{* \mu} \partial_{\mu} \mathcal{P} \bar{\mathcal{D}}-\mathcal{D} \partial_{\mu} \mathcal{P} \bar{\mathcal{D}}^{* \mu}\right), \nonumber\\
\mathcal{L}_{\mathcal{D}^{*} \mathcal{D} \mathcal{V}}&=&-2 f_{\mathcal{D}^{*} \mathcal{D} \mathcal{V}} \epsilon_{\mu v \alpha \beta}\left(\partial^{\mu} \mathcal{V}^{v}\right)_{j}^{i}\left(\mathcal{D}_{i}^{\dagger} \stackrel{\leftrightarrow}{\partial}_{\alpha} \mathcal{D}^{* \beta j}\right.\nonumber\\&-&\left.\mathcal{D}_{i}^{* \beta \dagger} \stackrel{\leftrightarrow}{\partial}_{\alpha} \mathcal{D}^{j}\right),\nonumber\\
\mathcal{L}_{\mathcal{D}^{*} \mathcal{D}^{*} \mathcal{P}}&=&\frac{1}{2} g_{\mathcal{D}^{*} \mathcal{D}^{*} \mathcal{P}} \varepsilon_{\mu \nu \alpha \beta} \mathcal{D}_{i}^{* \mu} \partial^{v} \mathcal{P}^{i j} \stackrel{\leftrightarrow}{\partial}^{\alpha} \mathcal{D}_{j}^{* \beta \dagger},\nonumber\\
\mathcal{L}_{\mathcal{D}^{*} \mathcal{D}^{*} \mathcal{V}}&=&i g_{\mathcal{D}^{*} \mathcal{D}^{*} \mathcal{V}} \mathcal{D}_{i}^{* \nu \dagger} \stackrel{\leftrightarrow}{\partial}_{\mu}\mathcal{D}_{\nu}^{*j}\left(\mathcal{V}^{\mu}\right)_{j}^{i}\nonumber\\&+&4 i f_{\mathcal{D}^{*} \mathcal{D}^{*} \mathcal{V}} \mathcal{D}_{i \mu}^{* \dagger}\left(\partial^{\mu} \mathcal{V}^{\nu}-\partial^{\nu} \mathcal{V}^{\mu}\right)_{j}^{i} \mathcal{D}_{\nu}^{* j}+\text { H.c.},\nonumber
\end{eqnarray}
\begin{eqnarray}
\mathcal{L}_{\mathcal{D}_{1} \mathcal{D}^{*} \mathcal{P}}&=& g_{\mathcal{D}_{1} \mathcal{D}^{*} \mathcal{P}}\left[3 \mathcal{D}_{1}^{\mu}\left(\partial_{\mu} \partial_{\nu} \mathcal{P}\right) \mathcal{D}^{* \nu \dagger}-\mathcal{D}_{1}^{\mu}\left(\partial^{\nu} \partial_{\nu} \mathcal{P}\right) \mathcal{D}_{\mu}^{* \dagger}\right] \nonumber\\&+&\text { H.c.},\nonumber\\
\mathcal{L}_{\mathcal{D}_{1}^{\prime} \mathcal{D}^{*} \mathcal{P}}&=&i g_{\mathcal{D}_{1}^{\prime} \mathcal{D}^{*} \mathcal{P}}\left(\mathcal{D}_{1}^{\prime \mu} \stackrel{\leftrightarrow}{\partial}_{\nu} \mathcal{D}_{\mu}^{* \dagger}\right) \partial^{\nu} \mathcal{P}+\text { H.c.},
\label{Eq:Lag1}
\end{eqnarray}
where $\mathcal{D}^{(*)\dagger}=(\bar{D}^{(*)0},D^{(*)-},D_{s}^{(*)-})$. The symbols $\mathcal{V}$ and $\mathcal{P}$ are the matrixes form of vector nonet and pseudoscalar nonet, respectively, their concrete form are,
\begin{eqnarray}
\mathcal{V}=
\begin{pmatrix}
\frac{1}{\sqrt2}(\rho^{0}+\omega)&\rho^{+}&K^{*+}\\
\rho^{-}&
\frac{1}{\sqrt2}(-\rho^{0}+\omega)&K^{*0}\\
K^{*-}&\bar{K}^{*0}&\phi\\ \nonumber
\end{pmatrix},
\end{eqnarray}
\begin{eqnarray}
\mathcal{P}=
\begin{pmatrix}
\frac{\pi^{0}}{\sqrt{2}}+\alpha\eta+\beta\eta^{\prime}&\pi^{+}&K^{+}\\
\pi^{-}&-\frac{\pi^{0}}{\sqrt{2}}++\alpha\eta+\beta\eta^{\prime}&K^{0}\\
K^{-}&\bar{K}^{0}&\gamma\eta+\delta\eta^{\prime}\\
\end{pmatrix},
\end{eqnarray}
where $\alpha$, $\beta$, $\gamma$ and~$\delta$ are the parameters related to the mixing angle, which are,
\begin{eqnarray}
\alpha &=&\frac{\mathrm{cos}\theta-\sqrt{2}\mathrm{sin}\theta}{\sqrt{2}}, \  \ \ \ \ \beta=\frac{\mathrm{sin}\theta+\sqrt{2}\mathrm{cos}\theta}{\sqrt{6}},\nonumber\\ \gamma &=&\frac{-2\mathrm{cos}\theta-\sqrt{2}\mathrm{sin}\theta}{\sqrt{6}}, \ \ \ \ \delta=\frac{-2\mathrm{sin}\theta+\sqrt{2}\mathrm{cos}\theta}{\sqrt{6}	},
\end{eqnarray}
where the mixing angle $\theta$ is determined to be $19.1^{\circ}$~\cite{MARK-III:1988crp,DM2:1988bfq}.

Considering SU(3) symmetry, the effective Lagrangians between light pseudoscalar mesons and vector mesons can be~\cite{Xiao:2020ltm,Chen:2011cj,Oh:2000qr,Haglin:2000ar},
\begin{eqnarray}
\mathcal{L}_{K^{*} K P}&=&-i g_{K^{*} K P}\left(\bar{K} \partial^{\mu} P-\partial^{\mu} \bar{K}P\right) K_{\mu}^{*}+\text { H.c.}, \nonumber\\
\mathcal{L}_{K^{*} K^{*} P}&=&-g_{K^{*} K^{*} P} \epsilon^{\mu \nu \alpha \beta} \partial_{\alpha} \bar{K}_{\beta}^{*} P \partial_{\mu} K_{\nu}^{*},\nonumber\\
\mathcal{L}_{K^{*} K V}&=&-g_{K^{*} K V} \epsilon^{\eta \tau \rho \sigma} \partial_{\rho} \bar{K}_{\sigma}^{*} \partial_{\eta} V_{\tau} K+\text { H.c.},\nonumber\\
\mathcal{L}_{K^{*} K^{*} V}&=&-i g_{K^{*} K^{*} V}\left[\left(\partial^{\mu} \bar{K}_{\nu}^{*} V^{\nu}-\bar{K}_{\nu}^{*} \partial^{\mu} V^{\nu}\right) K_{\mu}^{*}\right.\nonumber\\
&+&\bar{K}_{\mu}^{*}\left(\partial^{\mu} V^{\nu} K_{\nu}^{*}-V^{\nu} \partial^{\mu} K_{\nu}^{*}\right) \nonumber\\
&\times&\left.+\left(\bar{K}_{\nu}^{*} V^{\mu} \partial_{\mu} K^{* \nu}-\partial_{\mu} \bar{K}_{\nu}^{*} V^{\mu} K^{* \nu}\right)\right],
\label{Eq:Lag2}
\end{eqnarray}
where $P$ refers to the triplet of $\pi$, $\eta$ and $\eta^{\prime}$ from pseudoscalar nonet, and the vector meson $V$ stands for $\rho$ triplet and $\omega$ from vector nonet. The doublet $K^{(*)}$ is $K^{(*)}=(K^{(*)-},\bar{K}^{(*)0})$. The relevant coupling constants in Eqs.~(\ref{Eq:Lag1}) and (\ref{Eq:Lag2}) will be discussed in the following section.

\subsection{Two body decay process}

According to the effective Lagrangians listed above, we can abtain the amplitude for $T_{c\bar{s}0}^{0}\to D^{0}K^{0}$,~$D_{s}^{+}\pi^{-}$,~$D_{s}^{*+}\rho^{-}$,~$D_{s1}^{(\prime)+}\pi^{-}$ corresponding to diagrams in Fig~\ref{Fig:Tri2}, which are,
\begin{eqnarray}
i\mathcal{M}_{a}&=&i^{3}\int \frac{d^{4}q}{(2\pi)^4}\left[g_{T}\tilde{\Phi}(-p_{12}^{2},\Lambda^{2}_{T}) g ^{\phi\tau}\right]\nonumber\\
&\times &\left[ig_{D^{*}DP}\left(iq^{\mu}\right)g^{\mu\delta}\right]\left[-ig_{K^{*}KP}\left(ip_{4}^{\nu}+iq^{\nu}\right)g^{\nu a}\right]\nonumber\\&\times&\frac{-g^{\phi\delta}+p_{1}^{\phi}p_{1}^{\delta}/m_{1}^{2}}{p_{1}^{2}-m_{1}^{2}}\frac{-g^{\tau a}+p_{2}^{\tau}p_{2}^{a}/m_{2}^{2}}{p_{2}^{2}-m_{2}^{2}}\frac{1}{q^{2}-m_{q}^{2}}\nonumber\\&\times&\mathcal{F}^{2}\left(m_{q},\Lambda \right),\nonumber
\end{eqnarray}
\begin{eqnarray}
i\mathcal{M}_{b}&=&i^{3}\int \frac{d^{4}q}{(2\pi)^4}\left[g_{T}\tilde{\Phi}(-p_{12}^{2},\Lambda^{2}_{T}) g ^{\phi\tau}\right]\nonumber\\&\times&\left[-2f_{D^{*}DV}\epsilon_{\mu\nu\alpha\beta}(iq^{\mu})g^{\nu\lambda}(-ip_{1}^{\alpha}-ip_{3}^{\alpha})g^{\beta\delta}\right]\nonumber\\&\times&\left[-g_{K^{*}K^{*}P}\epsilon_{\omega\theta\rho\sigma}(-ip_{2}^{\omega})(-iq^{\rho})g^{\theta\xi}g^{\sigma a}\right]\nonumber\\&\times&\frac{-g^{\phi\delta}+p_{1}^{\phi}p_{1}^{\delta}/m_{1}^{2}}{p_{1}^{2}-m_{1}^{2}}\frac{-g^{\tau a}+p_{2}^{\tau}p_{2}^{a}/m_{2}^{2}}{p_{2}^{2}-m_{2}^{2}}\nonumber\\&\times&\frac{-g^{\lambda\xi}+q^{\lambda}q^{\xi}/m_{q}^{2}}{q^{2}-m_{q}^{2}}\mathcal{F}^{2}(m_{q},\Lambda),\nonumber
\end{eqnarray}
\begin{eqnarray}
i\mathcal{M}_{c}&=&i^{3}\int \frac{d^{4}q}{(2\pi)^4}\left[g_{T}\tilde{\Phi}(-p_{12}^{2},\Lambda^{2}_{T}) g ^{\phi\tau}\right]\nonumber\\&\times&\left[-ig_{D^{*}DP}(iq^{\mu})g^{\mu\delta}\right]\left[-ig_{K^{*}KP}(-iq^{\nu}-ip_{4}^{\nu})g^{\nu a}\right]\nonumber\\&\times&\frac{-g^{\phi\delta}+p_{1}^{\phi}p_{1}^{\delta}/m_{1}^{2}}{p_{1}^{2}-m_{1}^{2}}\frac{-g^{\tau a}+p_{2}^{\tau}p_{2}^{a}/m_{2}^{2}}{p_{2}^{2}-m_{2}^{2}}\frac{1}{q^{2}-m_{q}^{2}}\nonumber\\&\times&\mathcal{F}^{2}(m_{q},\Lambda),\nonumber
\end{eqnarray}
\begin{eqnarray}
i\mathcal{M}_{d}&=&i^{3}\int \frac{d^{4}q}{(2\pi)^4}\left[g_{T}\tilde{\Phi}(-p_{12}^{2},\Lambda^{2}_{T}) g ^{\phi\tau}\right]\nonumber\\&\times&\left[2f_{D^{*}DV}\epsilon_{\mu\nu\alpha\beta}(iq^{\mu})g^{\nu\lambda}(ip_{3}^{\alpha}+ip_{1}^{\alpha})g^{\beta\delta}\right]\nonumber\\&\times&\left[g_{K^{*}KV}\epsilon_{\omega\theta\rho\sigma}(-iq)^{\omega}(-ip_{2})^{\rho}g^{\theta\xi}g^{\sigma a}\right]\nonumber\\&\times&\frac{-g^{\phi\delta}+p_{1}^{\phi}p_{1}^{\delta}/m_{1}^{2}}{p_{1}^{2}-m_{1}^{2}}\frac{-g^{\tau a}+p_{2}^{\tau}p_{2}^{a}/m_{2}^{2}}{p_{2}^{2}-m_{2}^{2}}\nonumber\\&\times&\frac{-g^{\lambda\xi}+q^{\lambda}q^{\xi}/m_{q}^{2}}{q^{2}-m_{q}^{2}}\mathcal{F}^{2}(m_{q},\Lambda),\nonumber
\end{eqnarray}
\begin{eqnarray}
i\mathcal{M}_{e}&=&i^{3}\int \frac{d^{4}q}{(2\pi)^4}\left[g_{T}\tilde{\Phi}(-p_{12}^{2},\Lambda_{T}^{2}) g ^{\phi\tau}\right]\nonumber\\&\times&\left[-\frac{1}{2}g_{D^{*}D^{*}P}\epsilon_{\mu\nu\alpha\beta}g^{\mu\delta}(iq)^{\nu}(ip_{3}^{\alpha}+ip_{1}^{\alpha})g^{\beta\theta}\varepsilon^{b}({p_{3}})\right]\nonumber\\&\times&\left[-g_{K^{*}KV}\epsilon_{\eta mn\sigma}(-ip_{2})^{n}g^{\sigma a}(ip_{4})^{\eta}g^{m\rho}\varepsilon^{c}({p_{4}})\right]\nonumber\\&\times&\frac{-g^{\phi\delta}+p_{1}^{\phi}p_{1}^{\delta}/m_{1}^{2}}{p_{1}^{2}-m_{1}^{2}}\frac{-g^{\tau a}+p_{2}^{\tau}p_{2}^{a}/m_{2}^{2}}{p_{2}^{2}-m_{2}^{2}}\frac{1}{q^{2}-m_{q}^{2}}\nonumber\\&\times&\mathcal{F}^{2}(m_{q},\Lambda),\nonumber
\end{eqnarray}
\begin{eqnarray}
i\mathcal{M}_{f}&=&i^{3}\int \frac{d^{4}q}{(2\pi)^4}\left[g_{T}\tilde{\Phi}(-p_{12}^{2},\Lambda^{2}_{T}) g ^{\phi\tau}\right]\nonumber\\&\times&\left[ig_{D^{*}D^{*}V}(-ip_{1}^{\mu}-ip_{3}^{\mu})g^{\mu\lambda}g^{\nu\delta}g^{\nu\theta}\right.\nonumber\\&+&\left.4if_{D^{*}D^{*}V}g^{\mu\theta}(iq^{\mu}g^{\nu\lambda}-iq_{\nu}g^{\mu\lambda})g^{\nu\delta}\varepsilon^{b}({p_{3}})\right]\nonumber\\&\times&\left[-ig_{K^{*}K^{*}V}(((-ip_{2})^{\mu}g^{\nu a}g^{\nu\rho}-(ip_{4})^{\mu}g^{\nu a}g^{\nu\rho})g^{\mu\xi}\right.\nonumber\\&+&g^{\mu a}((ip_{4}^{\mu}g^{\nu\rho}g^{\nu\xi}-g^{\nu\rho}(-iq)^{\mu}g^{\nu\xi})\nonumber\\&+&\left.g^{\nu a}g^{\mu\rho}(-iq)^{\mu}g^{\nu\xi}-(-ip_{2})^{\mu}g^{\nu a}g^{\mu\rho}g^{\nu\xi}\varepsilon^{c}({p_{4}})\right]\nonumber\\&\times&\frac{-g^{\phi\delta}+p_{1}^{\phi}p_{1}^{\delta}/m_{1}^{2}}{p_{1}^{2}-m_{1}^{2}}\frac{-g^{\tau a}+p_{2}^{\tau}p_{2}^{a}/m_{2}^{2}}{p_{2}^{2}-m_{2}^{2}}\nonumber\\&\times&\frac{-g^{\lambda\xi}+q^{\lambda}q^{\xi}/m_{q}^{2}}{q^{2}-m_{q}^{2}}\mathcal{F}^{2}(m_{q},\Lambda),\nonumber
\end{eqnarray}
\begin{eqnarray}
i\mathcal{M}_{g}&=&i^{3}\int \frac{d^{4}q}{(2\pi)^4}\left[g_{T}\tilde{\Phi}(-p_{12}^{2},\Lambda^{2}_{T}) g ^{\phi\tau}\right]\nonumber\\&\times&\left[-ig_{K^{*}KP}i(p_{3}^{\mu}-q^{\mu})g^{\mu\delta}\right]\left[g_{D_{1}D^{*}P}(3g^{\omega\rho}(-iq)^{\omega}\right.\nonumber\\&\times&\left.(-iq^{\nu})g^{\nu a}-g^{\omega\rho}(-iq)^{\nu}(-iq)^{\nu}g^{\omega a})\varepsilon^{t}({p_{4}})\right]\nonumber\\&\times&\frac{-g^{\phi\delta}+p_{1}^{\phi}p_{1}^{\delta}/m_{1}^{2}}{p_{1}^{2}-m_{1}^{2}}\frac{-g^{\tau a}+p_{2}^{\tau}p_{2}^{a}/m_{2}^{2}}{p_{2}^{2}-m_{2}^{2}}\nonumber\\&\times&\frac{1}{q^{2}-m_{q}^{2}}\mathcal{F}^{2}(m_{q},\Lambda),\nonumber
\end{eqnarray}
\begin{eqnarray}
i\mathcal{M}_{h}&=&i^{3}\int\frac {d^{4}q}{(2\pi)^4}\left[g_{T}\tilde{\Phi}(-p_{12}^{2},\Lambda^{2}_{T}) g ^{\phi\tau}\right]\nonumber\\&\times&\left[-ig_{K^{*}KP}i(p_{3}^{\mu}-q^{\mu})g^{\mu\delta}\right]\nonumber\\&\times&\left[g_{D_{1}^{\prime}D^{*}P}(-ip_{2}^{\nu}-ip_{4}^{\nu})g^{\omega a}g^{\omega\rho}(-iq)^{\nu}\varepsilon^{t}({p_{4}})\right]\nonumber\\&\times&\frac{-g^{\phi\delta}+p_{1}^{\phi}p_{1}^{\delta}/m_{1}^{2}}{p_{1}^{2}-m_{1}^{2}}\frac{-g^{\tau a}+p_{2}^{\tau}p_{2}^{a}/m_{2}^{2}}{p_{2}^{2}-m_{2}^{2}}\nonumber\\&\times&\frac{1}{q^{2}-m_{q}^{2}}\frac{1}{q^{2}-m_{q}^{2}}\mathcal{F}^{2}(m_{q},\Lambda).
\end{eqnarray}
In the above amplitudes, we introduce a form factor in monopole form to represent the inner structure and off-shell effect of the exchanging mesons, which is,
\begin{eqnarray}
\mathcal{F}\left(m_{q},\Lambda\right)=\frac{m_{q}^{2}-\Lambda^{2}}{q^2-\Lambda^{2}},
\end{eqnarray}
where $\Lambda$ is of order of 1 GeV.

The total amplitudes of $T_{c\bar{s}0}^{0}\to D_{s}^{+}\pi^{-}$, $D^{0}K^{0}$, $D^{*+}_{s}\rho^{-}$, $D_{s1}^{+}\pi^{-}$ and $D_{s1}^{\prime+}\pi^{-}$ are,
\begin{eqnarray}
\mathcal{M}_{T_{c\bar{s}0}^{0}\to D_{s}^{+}\pi^{-}}&=&\mathcal{M}_{a}+\mathcal{M}_{b},\nonumber\\
\mathcal{M}_{T_{c\bar{s}0}^{0}\to D^{0}\pi^{0}}&=&\mathcal{M}_{c}^{\pi^{0}}+\mathcal{M}_{c}^{\eta}+\mathcal{M}_{c}^{\eta^{\prime}}+\mathcal{M}_{d}^{\rho^{0}}+\mathcal{M}_{d}^{\omega},\nonumber\\
\mathcal{M}_{T_{c\bar{s}0}^{0}\to D^{*+}_{s}\rho^{-}}&=&\mathcal{M}_{e}+\mathcal{M}_{f},\nonumber\\
\mathcal{M}_{T_{c\bar{s}0}^{0}\to D_{s1}^{+}\pi^{-}}&=&\mathcal{M}_{g},\nonumber\\
\mathcal{M}_{T_{c\bar{s}0}^{0}\to D_{s1}^{\prime+}\pi^{-}}&=&\mathcal{M}_{h},\label{Eq:10}
\end{eqnarray}
respectively.

With the total amplitudes defined in Eq.~(\ref{Eq:10}), one can estimate the  partial width of the above decay processes by 
\begin{eqnarray}
\Gamma_{T_{c\bar{s}0}^{0}\to ...}&=&\frac{1}{8\pi}\frac{|\vec{p}|}{m_{T_{c\bar{s}0}^{0}}^{2}}\left|\ \overline{\mathcal{M}_{T_{c\bar{s}0}^{0}\to ...}}\ \right|^{2}.\label{Eq:x}
\end{eqnarray}
where $|\vec{p}|$ is the momentum of the final state in the initial state rest frame.

\begin{figure}[t]
  \centering
  \includegraphics[width=6cm]{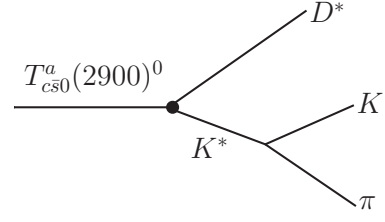}
\caption{The three body decay of the $T_{c\bar{s}0}^{0}$.}\label{Fig:Tri3}
\end{figure}

\subsection{Three body decay process}
\label{sec:Sec4}
In addition to the two-body decays, we also considered the contribution of the possible three-body decay process. The dominant three-body decay should be $T_{c\bar{s}0}^0 \to D^{\ast 0} K^{\ast0} \to D^{\ast 0} K \pi$, which is shown in Fig~\ref{Fig:Tri3}. The corresponding amplitude of the three-body decay is,
\begin{eqnarray}
\mathcal{M}&=&\left[g^{\phi\tau}\epsilon^{\phi}(p)\tilde{\Phi}(-p_{12}^{2},\Lambda^{2}_{T})\right]\left[ig_{K^{*}KP}g^{\mu\lambda}i\left(p_{3}^{\mu}-p_{2}^{\mu}\right)\right]\nonumber\\&\times&\frac{-g^{\tau\lambda}+q^{\lambda}q^{\tau}/m_{K^{*}}^{2}}{p_{2}^{2}-m_{K^{*}}^{2}+im_{K^{*}}\Gamma_{K^{*}}},
\end{eqnarray} 
and then the partial width is,
\begin{eqnarray}
d \Gamma =\frac{1}{(2 \pi)^{3}} \frac{1}{32 M^{3}} \overline{|\mathcal{M}|^{2}} d m_{12}^{2} d m_{23}^{2}.
\end{eqnarray}

\begin{figure}[t]
  \centering
  \includegraphics[width=8.5cm]{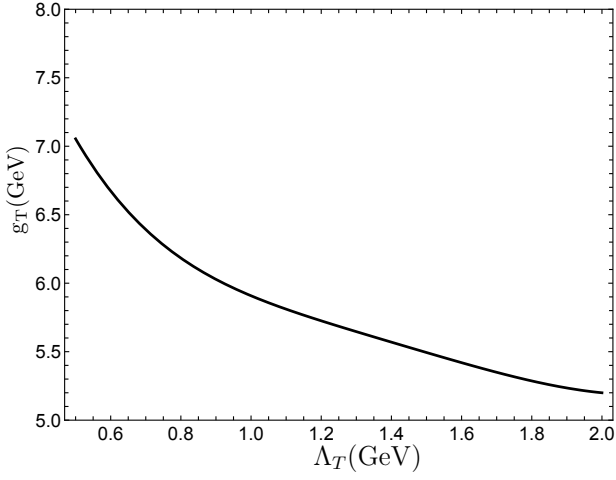}
  \caption{The coupling constant $g_{T}$ depending on model parameter $\Lambda_{T}$.}\label{Fig:Tri4}
\end{figure}

\section{Numerical results and discussions}
\label{sec:Sec4}
Before we estimate the widths of the considered channels, the relevant coupling constants should be clarified. Considering the heavy quark limit and chiral symmetry, the  coupling constants between the light mesons and charmed meson pairs are~\cite{Chen:2019asm,Liu:2011xc,Isola:2003fh,Falk:1992cx,Liu:2020ruo},
\begin{eqnarray}
g_{\mathcal{D}^{*}\mathcal{D}\mathcal{P}}&=&\frac{2g}{f_{\pi}}\sqrt{m_{\mathcal{D}}m_{\mathcal{D}}^{*}}=g_{\mathcal{D}^{*}\mathcal{D}^{*}\mathcal{P}}\sqrt{m_{\mathcal{D}}m_{\mathcal{D}^{*}}},\nonumber\\
f_{\mathcal{D}^{*}\mathcal{D}\mathcal{V}}&=&\frac{\lambda g_{V}}{\sqrt{2}}=\frac{f_{\mathcal{D}^{*}\mathcal{D}^{*}\mathcal{V}}}{m_{\mathcal{D}^{*}}},\nonumber\\
g_{\mathcal{D}^{*}\mathcal{D}^{*}\mathcal{V}}&=&\frac{\beta_{0}g_{V}}{\sqrt{2}},\nonumber\\
g_{\mathcal{D}_{1}\mathcal{D}^{*}P}&=&-2\sqrt{\frac{2}{3}}\frac{h^{\prime}}{\Lambda_{\chi}f_{\pi}}\sqrt{m_{\mathcal{D}_{1}}m_{\mathcal{D}^{*}}},\nonumber\\
g_{\mathcal{D}_{1}^{\prime}\mathcal{D}^{*}P}&=&-\frac{h}{f_{\pi}},
\end{eqnarray}
where $g=0.59$~\cite{Isola:2003fh} is determined by the measured width of $\Gamma(D^{*}\to D\pi)$. The values of the other parameters are $g_{V}=m_{\rho}/f_{\pi}$, $f_{\pi}=132 \mathrm{MeV}$, $\Lambda_{\chi}=1\mathrm{GeV}$, $\beta_{0}=0.9$ and $\lambda=0.56{\mathrm{GeV}}^{-1}$~\cite{Casalbuoni:1996pg}. The gauge couplings $h$ and $h^{\prime}$ are estimated to be $h=0.56\pm0.04$ and $h^{\prime}=0.43\pm0.01$~\cite{Ding:2008gr,Colangelo:2012xi}.

The coupling constants relevant to the light mesons are~\cite{Xiao:2020ltm,Chen:2011cj},
\begin{eqnarray}
g_{K^{*}KP}&=&g_{K^{*}K^{*}V}=\frac{1}{4}g,\nonumber\\
g_{K^{*}KV}&=&g_{K^{*}K^{*}P}=\frac{1}{4} \frac{g^{2} N_{c}}{16 \pi^{2} f_{\pi}},
\end{eqnarray}
where the parameter $g=0.78$ are determined via measured width of the process $K^{*}\to K\pi$. $N_{c}=3$ is the color degree of freedom.

Moreover, the coupling constant of $T_{c\bar{s}0}^{0}$ to its components, $g_{T}$, can be estimated by the compositeness condition as given in Eq.~(\ref{Eq:3}). The phenomenological parameter $\Lambda_{T}$ should be of the order of $1~\mathrm{GeV}$. In the present work, we varies $\Lambda_{T}$ in a sizable range from $0.5$ to $2.0~\mathrm{GeV}$. The numerical results of the $g_{T}$ depending on the parameter $\Lambda_T$ is presented in Fig.~\ref{Fig:Tri4}. In the considered parameter range, we can find the coupling constant $g_{T}$ decreases from $7.05$ to $5.20~\mathrm{GeV}$.

\begin{figure}[t]
 \centering
 \includegraphics[width=8.5cm]{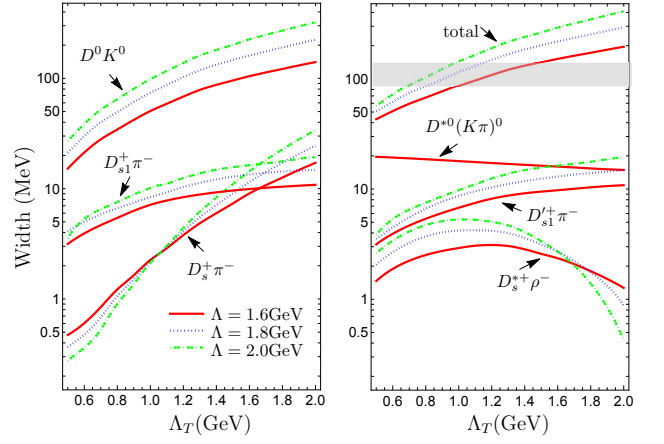}
  \caption{The partial widths of $T_{c\bar{s}0}^{0}\to D^{0}K^{0}$,~$D_{s}^{+}\pi^{-}$,~$D_{s}^{*+}\rho^{-}$,~$D_{s1}^{(\prime)+}\pi^{-}, D^{\ast 0} (K\pi)^0$ depending on the model parameter $\Lambda_T$ and $\Lambda$. The total width are the sum of the partial widths of the considered channels. The grey band are the width of $T_{c\bar{s}0}^0$ reported by LHCb Collaboratio~\cite{LHCb:New1, LHCb:New2}}\label{Fig:Tri5}
\end{figure}

With the above preparations, we can estimate the partial widths of $T_{c\bar{s}0}^{0}\to D^{0}K^{0}$,~$D_{s}^{+}\pi^{-}$,~$D_{s}^{*+}\rho^{-}$,~$D_{s1}^{(\prime)+}\pi^{-}, D^{\ast 0} (K\pi)^0$ depending on the model parameters $\Lambda_T$ and $\Lambda$, which were introduced by the correlation function of molecular state and the form factor in the amplitudes. These two parameters are all of order of 1 GeV. In the present estimations, we varies $\Lambda_T$ from 0.5 GeV to 2 GeV and take several typical values of $\Lambda$, which is 1, 1.5 and 2 GeV, respectively. In Fig.~\ref{Fig:Tri5}, we present our estimations of the widths of the considered processes. With the assumption that $T_{c\bar{s}0}^0$ dominantly decay into these final states, we can compare the sum of partial widths of the considered channels with the measured width, which is $(119\pm 26 \pm 12)$ MeV. From Fig.~\ref{Fig:Tri5}, we can find in the considered parameters space, our estimations of the total width can overlap with the experimental measurement from the LHCb Collaboration, in particular, the determined $\Lambda_T$ range are $1.03-1.56$, $0.83-1.19$, and $0.71-1.01~\mathrm{GeV}$ for $\Lambda=1.6$, $1.8$, and $2.0~\mathrm{GeV}$, respectively. In these parameter ranges the partial widths of the considered channels are presented in Table~\ref{Tab:Res}. From the table, one can find the branching ratio of the observed channel $D_s^+ \pi^-$ is $(0.65 \sim 5.38)\% $, which should be sizable to be observed experimentally. 

\begin{figure}[t]
 \centering
 \includegraphics[width=8.5cm]{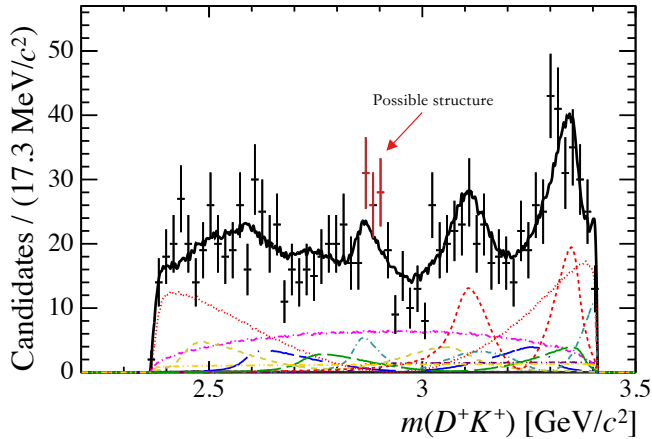}
  \caption{The $D^+K^+$ invariant mass distributions of the process $B^+\to D^+ D^- K^+$ reported by the LHCb Collaboration~\cite{LHCb:2020bls,LHCb:2020pxc}, where one can find a structure around 2.9 GeV, which may correspond to the contributions of $T_{c\bar{s}0}^{++}$.}\label{Fig:Tcsbarpp}
\end{figure}

Moreover, our estimations indicate that $T_{c\bar{s}0}^0$ dominantly decays into $D^0 K^0$. Considering the isospin symmetry, $D^+ K^+$ should be the dominant decay channel of $T_{c\bar{s}0}^{++}$. In Refs.~\cite{LHCb:2020bls,LHCb:2020pxc}, the LHCb Collaboration have measured the decay process $B^+ \to D^+ D^- K^+$, where two new structure $X_0(2900)$ and $X_1(2900)$ were observed in the $D^- K^+$ invariant mass distributions. Moreover, the LHCb Collaboration also reported their measurements of the $D^+K^+$ invariant mass distributions as shown in Fig.~\ref{Fig:Tcsbarpp}. From the figure one can find the fitted curve can not well describe the experimental data around 2.9 GeV, which indicates that there should exist an addition resonance. Since $T_{c\bar{s}0}^{++}$ dominantly decays into $D^+ K^+$ and its mass also well conform the one of the additional resonance, we believe that this structure may comes form the contribution of $T_{c\bar{s}0}^{++}$, which could be tested by further experimental analysis by LHCb Collaboration.

\begin{table}[t]
\begin{center}
\caption{The predicted partial widths of the considered  channels.}\label{Tab:Res}
\renewcommand\arraystretch{1.5}
  \setlength{\tabcolsep}{8mm}{
\begin{tabular}{ccc}
  \toprule[1pt]
Channel~&~width (MeV)\\
  \midrule[1pt]
   $T_{c\bar{s}0}^{0}\to D^{0}K^{0}$~&~$52.6-101.7$\\
 $T_{c\bar{s}0}^{0}\to D_{s}^{+}\pi^{-}$~&~$0.55-8.35$\\
 $T_{c\bar{s}0}^{0}\to D_{s}^{*+}\rho^{-}$~&~$2.96-5.3$\\
 $T_{c\bar{s}0}^{0}\to D_{s1}^{+}\pi^{-}$~&~$6.63-10.29$\\
 $T_{c\bar{s}0}^{0}\to D_{s1}^{\prime+}\pi^{-}$~&~$6.63-10.3$\\
$T_{c\bar{s}0}^{0}\to D^{*0}(K\pi)^{0}$~&~$16.11-18.96$\\
  \bottomrule[1pt]
\end{tabular}}
\end{center}
\end{table}

\section{Summary}
\label{sec:Sec5}

Very recently, two new resonances $T_{c\bar{s}0}^{0/++}$ were reported in $D_{s}\pi$ invariant mass spectrum of the processes $B^0 \to \bar{D}^0 D_s^+ \pi^-$ and $B^+ \to D^- D_s^+ \pi^+$ by the LHCb Collaboration. These two states are two of the isospin triplet, and the most possible $I(J^P)$ quantum numbers are $1(0^+)$. The observed mass of $T_{c\bar{s}0}$ is very close to the threshold of $D^\ast K^\ast$, which indicate that $T_{c\bar{s}0}$ could be a good candidate of hadronic molecular state composed of $D^\ast K^\ast$. 

In the present work, we investigate the decay behavior of $T_{c\bar{s}0}^0$ in the $D^\ast K^\ast$ molecular scenario by using an effective Lagrangian approach. Six possible dominant decay channels have been considered, which are $T_{c\bar{s}0}^{0}\to D^{0}K^{0}$,~$D_{s}^{+}\pi^{-}$,~$D_{s}^{*+}\rho^{-}$,~$D_{s1}^{(\prime)+}\pi^{-}$ and $D^{\ast 0} (K\pi)^0$. Our estimations indicate that the branching ratio of $T_{c\bar{s}0}^0\to D_s^+ \pi^-$ can reach up to $5.38\%$, which should be sizable to be observed.

Moreover, our estimations indicate that $T_{c\bar{s}0}^0$ dominantly decays into $D^0K^0$. Considering the isospin symmetry, one can expect that $D^+ K^+$ is the dominant decay mode of $T_{c\bar{s}0}^{++}$. By checking the $D^+ K^+$ invariant mass distributions of the process $B^+ \to D^+ D^- K^+$, we find the fitted curve can not describe the experimental data well around 2.9 GeV, which may indicate the signal of $T_{c\bar{s}0}^{++}$ in the process $B^+ \to D^+ D^- K^+$.

\bigskip
\noindent
\begin{center}
	{\bf ACKNOWLEDGEMENTS}\\
\end{center}
This work is supported by the National Natural Science Foundation of China under the Grant No.  12175037 and 11775050.

\end{document}